\documentclass{article}

\usepackage{amssymb}
\usepackage{amscd}
\usepackage{amsthm}

\newtheorem{thm}{Theorem}[section]
\newtheorem{prop}[thm]{Proposition}
\newtheorem{lem}[thm]{Lemma}
\newtheorem{cor}[thm]{Corollary}
\newtheorem{dfn}[thm]{Definition}
\newtheorem{rem}{Remark}

\newcommand{\Z}{\mathbb{Z}}
\newcommand{\R}{\mathbb{R}}
\newcommand{\C}{\mathbb{C}}
\newcommand{\T}{\mathbb{T}}
\newcommand{\g}{\mathfrak{g}}
\newcommand{\A}{{\mathcal A}}
\newcommand{\F}{{\mathcal F}}
\newcommand{\U}{{\mathcal U}}

\newcommand{\W}{{\mathcal W}}
\newcommand{\Cech}{$\check{\textrm{C}}$ech {}}

\renewcommand{\L}{{\mathcal L}}

\def\a#1#2{ \phi ^{#1}_{\sigma_ {#2}}} 
\def\M#1{ \mathrm{Map}({#1}, BG) }

\title{The formulation of the Chern-Simons action for \\
       general compact Lie groups using \\
       Deligne cohomology 
}

\author{Kiyonori Gomi
\thanks{The author's research is supported by Research Fellowship of the Japan Society for the Promotion of Science for Young Scientists.}}

\date{}

\begin{document}
\maketitle

%\marginpar{\footnotesize This manuscript is typeset by \LaTeXe with 
%packages amssymb, amscd and amsthm.  }

\begin{abstract}
We formulate the Chern-Simons action for any compact Lie group using Deligne cohomology. This action is defined as a certain function on the space of smooth maps from the underlying 3-manifold to the classifying space for principal bundles. If the 3-manifold is closed, the action is a ${\bf C}^*$-valued function. If the 3-manifold is not closed, then the action is a section of a Hermitian line bundle associated with the Riemann surface which appears as the boundary. 
\end{abstract}

%%%%%%%%%%%%%%%%%%%%%%%%%%%%%%%%%%%%%%%%%%%

\section{Introduction}

For a connection $A$ on a principal $SU(2)$-bundle over a compact oriented 3-manifold $M$, the \textit{Chern-Simons action functional (CS action)} \cite{W,F1} is
\begin{eqnarray*}
S(A) = \frac{1}{8\pi^2} \int_M 
Tr(A \wedge dA + \frac{2}{3} A \wedge A \wedge A).
\end{eqnarray*}
If $M$ has no boundary, then $e^{2\pi\sqrt{-1}S(A)}$ is $\C^*$-valued, but if $M$ has boundary, then $e^{2\pi\sqrt{-1}S(A)}$ takes values in the fiber of a Hermitian line bundle associated with the boundary. This description can be applied to all connected, simply connected compact Lie groups because any principal bundle of such a structure group is topologically trivial on a 3-manifold.

In the case of general compact Lie groups, Dijkgraaf and Witten \cite{D-W} defined an extension of the Chern-Simons action as follows. It is known that there exists a certain connection on the universal bundle $EG$ called the \textit{universal connection} \cite{N-R}. By means of the universal connection $A_u$ we can identify a smooth $G$-bundle equipped with a connection over $M$ with a smooth map from $M$ to the classifying space $BG$. Then we introduce \textit{differential characters} \cite{Chee-S}. A degree $p$ differential character $\alpha \in \hat{H}^p(X,\R/\Z)$ is by definition a homomorphism from the group of $p$-singular cycles of $X$ to $\R/\Z$ such that there exists a specified $(p+1)$-form $\omega$ and the relation $\alpha(\partial \tau) = \int_{\tau}\omega$ holds modulo $\Z$ for all $(p+1)$-singular chains. We define $\alpha_u \in \hat{H}^3(BG,\R/\Z)$ by the characteristic 4-form $\frac{1}{8\pi^2}Tr(F_u \wedge F_u)$, where $F_u$ is the curvature form of $A_u$. Then the CS action 
$$
S : \M{M} \rightarrow \R/\Z
$$
is defined by $S(\gamma)=\gamma^*\alpha_u(M)$. 

In this definition $M$ is treated as a 3-singular cycle. So this action is well defined only if the underlying 3-manifold $M$ has no boundary. Therefore we need some alternative tool in order to formulate the action when $M$ has a boundary. The purpose of this paper is to formulate the action using \textit{Deligne cohomology} \cite{B,B-M1,B-M2}. The Deligne cohomology $H^p(X,\F^q)$ is by definition the hypercohomology group of the complex of sheaves $\F^q :$
$$
\begin{CD}
\underline{\T}_X  @>\frac{1}{\sqrt{-1}}d \log>>
\underline{A}^1_X @>d>>
\cdots            @>d>>
\underline{A}^q_X,
\end{CD}
$$
where $\underline{\T}_X$ is the sheaf of unit circle valued functions on $X$ and $\underline{A}^p_X$ is the sheaf of $p$-forms on $X$. One of the reasons to introduce Deligne cohomology is that there is a natural isomorphism between the group of differential characters $\hat{H}^p(X,\R/\Z)$ and the Deligne cohomology group $H^p(X,\F^p)$. Another reason is that Gawedzki \cite{Ga} defined a topological term of the Wess-Zumino-Witten model for a general target space using Deligne cohomology. Applying his method to the 3-dimensional case, we can formulate the CS action as an analogy.

Using Deligne cohomology we define the CS action as follows. First we take a \Cech cocycle representation of a certain Deligne cohomology class $c_u \in H^3(BG, \F^3)$ which is uniquely determined by the characteristic 4-form on $BG$. Secondly we construct an open covering $\{ \W_Z \}$ of $\M{M}$, where $M$ is a compact 3-manifold. For each open set $\W_A$ a $\T$-valued function $\A_{M;A}$ is defined by integrating the \Cech cocycle. If $M$ is closed, these functions give rise to a global function which we call the CS action
$$
\A_M : \M{M} \rightarrow \T .
$$

If a 3-manifold is not closed, the local functions do not glue together to form a global function. But they give rise to a section of a certain Hermitian line bundle. For a closed Riemann surface $\Sigma$ the line bundle $\L_{\Sigma} \rightarrow \M{\Sigma}$ is constructed by transition functions with respect to an open covering. We call this line bundle the \textit{Chern-Simons line bundle}. The transition functions are defined by integrating the \Cech cocycle which represents $c_u$. The isomorphism class of this line bundle depends only on the Deligne cohomology class $c_u$. In the case of a 3-manifold $M$ with boundary, the local function $\A_{M;A}$ gives rise to a global section 
$$
\A_M : \M{M} \rightarrow r^*\L_{\partial M}  ,
$$
where $r$ is the restriction map to the boundary. We also call this section the CS action. The CS action defined here satisfies some properties that are considered to be the axioms of ``Classical Field Theory.'' In particular, it is compatible with the gluing operation of 3-manifolds.

\medskip

The organization of this paper is as follows. In Section \ref{sec_deligne_coh} we define Deligne cohomology and summarize basic properties. In Section \ref{sec_main} we define the Chern-Simons action for general compact Lie groups and the line bundle associated with a closed oriented Riemann surface. We also observe the fundamental properties of the action and the line bundle. 

\medskip

\textit{Acknowledgments.}
I would like to thank Prof. Kohno for useful comments on this subject.

%%%%%%%%%%%%%%%%%%%%%%%%%%%%%%%%%%%%%%%%%%%

\section{Deligne cohomology}
\label{sec_deligne_coh}

In this section we define \textit{Deligne cohomology} \cite{B,B-M1,B-M2} and observe its important properties.

\begin{dfn}
Let $\underline{A}^p_X$ be the sheaf of smooth $p$-forms on a smooth manifold $X$, and $\underline{\T}_X$ be the sheaf of unit circle valued functions on $X$. We define a complex of sheaves $\F^q$ by
$$
\begin{CD}
\underline{\T}_X  @>\frac{1}{\sqrt{-1}}d \log>>
\underline{A}^1_X @>d>>
\cdots            @>d>>
\underline{A}^q_X .
\end{CD}
$$
The hypercohomology group $H^p(X,\F^q)$ is called the Deligne cohomology group.
\end{dfn}

Usually we compute the Deligne cohomology by \Cech cohomology. Taking an open covering $\{U_i\}$ of $X$, we compute the \Cech cohomology $\check{H}^p(\{U_i\},\F^q)$. Then $H^p(X,\F^q)$ is obtained by the direct limit taken over the ordered set of open coverings of $X$, where the order is defined by the refinement of coverings. If we take a \textit{good} covering \cite{B-T}, then $\check{H}^p(\{U_i\},\F^q)$ is isomorphic to $H^p(X,\F^q)$. In this paper we use $D$ to denote the total differential operator on the \Cech double complex.

\begin{thm}[Brylinski \cite{B}] \label{thm1}
The Deligne cohomology $H^p(X,\F^p)$ fits in the following exact sequences
\begin{eqnarray*}
&
0 \longrightarrow H^p(X,\T)  \longrightarrow  H^p(X,\F^p) 
\stackrel{\delta_1}{\longrightarrow} A^{p+1}(X)_0 \longrightarrow 0 ,
& \\
&
0 \longrightarrow  A^p(X)/A^p(X)_0 \longrightarrow  H^p(X,\F^p) 
\stackrel{\delta_2}{\longrightarrow} H^{p+1}(X,\Z) \longrightarrow 0 , 
& \\
&
0 \longrightarrow H^p(X,\R)/H^p(X,\Z)\longrightarrow  H^p(X,\F^p) 
\stackrel{(\delta_1,\delta_2)}{\longrightarrow} 
R^{p+1}(X,\Z) \longrightarrow 0 ,
&
\end{eqnarray*}
where $A^k(X)_0$ is the group of integral $k$-forms on $X$. $R^k(X,\Z)$ is defined by
$$
\{ (\omega,u) \in A^k(X)_0 \times H^k(X,\Z) |
r(u) = [\omega], r : H^k(X,\Z) \rightarrow H^k(X,\R) \} .
$$
The homomorphisms $\delta_1$, $\delta_2$ are defined as follows. Let $(g, \omega^1, \cdots, \omega^p)$ be a \Cech cocycle representation of a Deligne cohomology class. The image under $\delta_1$ is just $\frac{1}{2\pi} d\omega^p$ glued together, and the image under $\delta_2$ is the cocycle $\frac{1}{\sqrt{-1}}\log \delta g$, where $\delta$ is the \Cech derivation.
\end{thm}

A trivial example of Deligne cohomology is $H^0(X,\F^0)$. We can easily see that $H^0(X,\F^0) \cong H^0(X,\underline{\T}_X)$. In this case the Deligne cohomology group is the group of $\T$-valued functions on $X$. As another example of Deligne cohomology, we explain the classification of Hermitian line bundles with Hermitian connection. The set of isomorphism classes of line bundles over $X$ has a group structure realized by tensor products of line bundles. It is well known that $H^2(X,\Z) \cong H^1(X,\underline{\T}_X)$ is isomorphic to the group of isomorphism classes of line bundles over $X$. If we endow a connection with a line bundle, we can classify the isomorphism classes of them in terms of the Deligne cohomology.

\begin{thm}[Brylinski \cite{B}] \label{thm2}
The group of isomorphism classes of Hermitian line bundles with Hermitian connection over $X$ is isomorphic to the Deligne cohomology group $H^1(X,\F^1)$.
\end{thm}

\begin{proof}
Let $(L,\nabla)$ be a Hermitian line bundle with a Hermitian connection over $X$. We can take an open covering $\{U_i\}$ such that there are local sections $\{ s_i : U_i \rightarrow L|_{U_i} \}$. We define the transition functions $\{ g_{ij} : U_i \cap U_j \rightarrow \C^* \}$ by $s_i = g_{ij} s_j$. Because $L$ is equipped with a Hermitian metric, $g_{ij}$ takes values in $\T$. By the Hermitian connection $\nabla$ we define the connection forms $\{ \alpha_i \}$ by $\nabla s_i = -\sqrt{-1}\alpha_i\otimes s_i$. Then we have a \Cech cochain $(g_{ij}, \alpha_i) \in \check{C}^1(\{U_i\},\F^1)$. This cochain is a cocycle because the following relations are satisfied:
\begin{eqnarray*}
g_{ij}g_{jk}g_{ki}  & = & 1 , \\
\alpha_i - \alpha_j  & =  & \frac{1}{\sqrt{-1}} d \log g_{ij} .
\end{eqnarray*}
If we take other local section $s_i'$, then we have the \Cech cocycle $(g'_{ij}, \alpha'_i)$ which is cohomologous to $(g_{ij}, \alpha_i)$. This construction of the \Cech cocycle gives a homomorphism from the isomorphism classes of Hermitian line bundles with Hermitian connection to $H^1(X,\F^1)$. If local sections $\{ s_i \}$ give $(g_{ij}, \alpha_i)$ whose cohomology class is trivial, then we have a global horizontal section of $L$. This implies that $(L, \nabla)$ is trivial and the homomorphism is injective. If we are given a cohomology class of $H^1(X,\F^1)$, then we express the class by a \Cech cocycle with respect to an open covering. Obviously the \Cech cocycle gives rise to a Hermitian line bundle with a Hermitian connection. Hence the homomorphism is surjective and the theorem is proved.
\end{proof}

If a Hermitian line bundle with a Hermitian connection is given by a Deligne cohomology class, the curvature form and the 1st Chern class correspond to the images under $-\delta_1, \delta_2$ of Theorem \ref{thm1} respectively.

\smallskip

There exists a natural isomorphism between the differential character group of Cheeger-Simons and the Deligne cohomology.

\begin{dfn}[Cheeger-Simons \cite{Chee-S}]
Let $X$ be a smooth manifold and $S_q(X)$ the group of smooth $q$-singular cycles of $X$. A differential character $\alpha$ of degree $p$ is defined as a homomorphism $\alpha : S_p(X) \rightarrow \R/\Z$ such that there exists a specific $(p+1)$-form $\omega$ satisfying $\alpha(\partial \tau) = \int_\tau \omega$ modulo $\Z$ for all $(p+1)$-singular chains of $X$. The group of such homomorphisms is called the degree $p$ differential character group and is denoted by $\hat{H}^p(X,\R/\Z)$.
\end{dfn}

\begin{thm}[Brylinski-McLaughlin \cite{B-M1}] \label{thm3}
There exists a natural isomorphism
$$
\hat{H}^p(X,\R/\Z) \cong H^p(X,\F^p) .
$$
\end{thm}

%%%%%%%%%%%%%%%%%%%%%%%%%%%%%%%%%%%%%%%%%%%

\section{The CS action for a general compact Lie group}
\label{sec_main}

In this section we observe the formulation of the Chern-Simons action functional using Deligne cohomology. Roughly speaking, (the exponential of $2\pi\sqrt{-1}$ times) the CS action is a $\C^*$-valued function on the set of principal bundles with connection. So we have to choose a bundle with connection before we define the action. Firstly we fix a compact Lie group $G$. It is well known that any smooth $G$-bundle $E$ over a manifold $M$ can be realized as the pull back of the universal bundle $EG$ by a smooth map $\gamma$ from $M$ to the classifying space $BG$. More generally, it is known by Narashimahn and Ramanan \cite{N-R} that there exists the so called \textit{universal connection} $A_u$ on $EG$ for any compact Lie group $G$, and any connection $A$ on $E$ can also be realized as the pull back of $A_u$ by the smooth map from $E$ to $EG$ covering $\gamma : M \rightarrow BG$. 
%Moreover two maps which induce isomorphic bundles with connection are homotopic.
This enables us to identify a smooth map from $M$ to $BG$ with a $G$-bundle equipped with a connection. We denote the space of smooth maps from $M$ to $BG$ by $\M{M}$. The map space has the compact-open topology.

We choose a Deligne cohomology class of $H^3(BG,\F^3)$. On $EG$ we have the universal connection $A_u$ which we fix throughout this paper. The standard argument of the characteristic classes implies that the 4-form $\frac{1}{8\pi^2}Tr(F_u \wedge F_u)$ is an integral 4-form on $BG$, where $F_u$ is the curvature form of $A_u$. It is well known that the odd cohomology groups of $BG$ vanish. By the help of Theorem \ref{thm1}, we can uniquely choose a Deligne cohomology class $c_u$ that corresponds to the integral 4-form above. If we take a good covering $\{ U_\alpha \}$ of $BG$, then we can express $c_u$ by a \Cech cocycle $( g_{\alpha_0 \alpha_1 \alpha_2 \alpha_3}, \omega^1_{\alpha_0 \alpha_1 \alpha_2}, \omega^2_{\alpha_0 \alpha_1}, \omega^3_{\alpha} ) \in \check{C}^3(\{U_\alpha\},\F^3)$ such that
$\frac{1}{2\pi} d\omega^3_\alpha = \frac{1}{8\pi^2}Tr(F_u \wedge F_u) |_{U_\alpha}$. The explicit formula of this cocycle is known by Brylinski and McLaughlin \cite{B-M1}.

\begin{dfn}  \label{dfn_open_set}
Let $\{ U_\alpha \}_{\alpha \in I}$ be an open covering of $BG$ and $K = \{\sigma_0, \sigma_1, \sigma_2, \sigma_3 \}$ a triangulation of a compact oriented smooth 3-manifold $M$, where $\sigma_p$ denotes the $p$-simplex. For the covering and the triangulation, we choose a map $\phi : K \rightarrow I$. We denote the image of $\sigma$ by $\phi_\sigma$. For a pair $A = \{K, \phi \}$, we define an open set of $\M{M}$ by
$$
\W_A = \{\gamma \in \M{M} |
\gamma(\sigma_p) \subset U_{\a{}{p}} (p=0,1,2,3) \} .
$$
All choices of $A = \{K, \phi \}$ give the covering $\{ \W_A \}$ of $\M{M}$.
\end{dfn}

\begin{lem}
(1) Let $\W_A$ be an open set with $A=\{K, \phi \}$, and $K'$ a subdivision of the triangulation $K$. We define an induced map $\phi': K' \rightarrow I$ as follows. For each simplex $\sigma' \in K'$ there exists a unique simplex $\sigma \in K$ including $\sigma'$ whose dimension is the lowest. The map is defined by $\phi'_{\sigma'} = \phi_\sigma$. For $A' = \{K', \phi' \}$ we have $\W_A$ = $\W_{A'}$ as sets of maps.

(2) Let $\W_{A_0}, \W_{A_1}$ be open sets with $A_0 = \{K_0, \phi^0 \}, A_1 = \{K_1, \phi^1 \}$. We take a common subdivision $\tilde{K}$ of $K_0, K_1$. To the subdivision we induce maps $\tilde{\phi}^0, \tilde{\phi}^1$ from $\phi^0, \phi^1$ respectively. We put $\tilde{A}_0 = \{\tilde{K}, \tilde{\phi}^0 \}, \tilde{A}_1 = \{\tilde{K}, \tilde{\phi}^1 \}$. If we define the intersection of $\W_{A_0}$ and $\W_{A_1}$ by
\begin{eqnarray*}
\W_{A_0} \cap \W_{A_1} & := & \W_{\tilde{A}_0} \cap \W_{\tilde{A}_1} \\
& = & \{\gamma \in \M{M} | \gamma(\sigma_p) 
\subset U_{\tilde{\phi}^0_{\sigma_p}} \cap U_{\tilde{\phi}^1_{\sigma_p}}
 (p=0,1,2,3) \} ,
\end{eqnarray*}
then the intersection is independent of the choice of the subdivisions. 
\end{lem}

\begin{rem}
For triangulations $K_1$ and $K_2$ such that $|K_1| = |K_2|$, there exists a common subdivision of them \cite{M}.
\end{rem}

\begin{proof}
We can directly verify (1) by the definition of the induced map $\phi': K' \rightarrow I$. Using (1) we can prove (2).
\end{proof}

For a triangulation $K = \{ \sigma_0, \ldots, \sigma_d \}$ of a $d$-dimensional manifold $X$, we define the set of \textit{flags} of simplices of $K$ by 
$$
F_K(p) = \{(\sigma_p, \sigma_{p+1}, \ldots,\sigma_d) | \sigma_i \in K, 
\sigma_i \subset \partial \sigma_{i+1}, i=p, \dots, d-1   \} .
$$
If $X$ is oriented, $\sigma_d$ has the orientation which is compatible with that of $X$. If a flag $(\sigma_p, \ldots,\sigma_d)$ is given, then the orientation of $\sigma_p$ is induced from that of $\sigma_d$. We perform the integration over a simplex along the induced orientation.

\begin{dfn}
Let $M$ be a compact oriented smooth 3-manifold, $\{U_\alpha\}$ a good covering of $BG$, and $(g, \omega^1, \omega^2, \omega^3)$ a \Cech cocycle representation of $c_u$. From the covering of $BG$ we have the open covering $\{ \W_A \}$ of $\M{M}$. For each open set $\W_A \neq \emptyset$ we define a function
\begin{eqnarray*}
\A_{M;A} : \W_A \rightarrow \T
\end{eqnarray*}
by the following
\begin{eqnarray*}
\A_{M;A}(\gamma) & = &
\exp \sqrt{-1} \left\{
\sum_{F_K(3)} 
 \int_{\sigma_3} \gamma^*
  \omega^3_{\a{}{3}}
-
\sum_{F_K(2)}
 \int_{\sigma_2} \gamma^*
  \omega^2_{\a{}{2} \a{}{3}}
\right.   \\
&   & \left.   
-
\sum_{F_K(1)}
 \int_{\sigma_1} \gamma^*
  \omega^1_{\a{}{1} \a{}{2} \a{}{3}}
\right\} \times
\prod_{F_K(0)}
 g_{\a{}{0} \a{}{1} \a{}{2} \a{}{3}}(\gamma(\sigma_0))  .
\end{eqnarray*}
\end{dfn}

\begin{lem}
Let $\W_A$ be an open set of $\M{M}$ with $A=\{ K,\phi \}$. For a subdivision $K'$ of $K$, we induce the map $\phi' : K' \rightarrow I$ and put $A'= \{ K', \phi' \}$. On $\W_A = \W_{A'}$ we have
$$
\A_{M;A}(\gamma) = \A_{M;A'}(\gamma) .
$$
\end{lem}

\begin{proof}
By the construction of $\phi'$ we have
\begin{eqnarray*}
\sum_{F_{K'}(3)} \int_{\sigma'_3} \gamma^* \omega^3_{\phi'_{\sigma'_3}}
& = &
\sum_{\sigma_3} \sum_{ {\sigma'_3} \atop {\sigma'_3 \subset \sigma_3} }
\int_{\sigma'_3} \gamma^* \omega^3_{\phi'_{\sigma'_3}}
=
\sum_{\sigma_3} \sum_{{\sigma'_3} \atop {\sigma'_3 \subset \sigma_3}}
\int_{\sigma'_3} \gamma^* \omega^3_{\phi_{\sigma_3}} \\
& = &
\sum_{F_{K}(3)} \int_{\sigma_3} \gamma^* \omega^3_{\phi_{\sigma_3}} .
\end{eqnarray*}
We can calculate the other terms using the property of the \Cech cocycle, i.e.\@ $\omega_{\ldots \alpha \ldots \alpha \ldots} = 0$. So the lemma is proved.
\end{proof}

\begin{lem} \label{lem_difference_of_action_1}
Over the intersection $\W_{A_0} \cap \W_{A_1} \neq \emptyset$ we have
\begin{eqnarray*}
&   &
\A_{M;A_1}(\gamma) \A_{M;A_0}(\gamma)^{-1} \\
&   & =
\exp \sqrt{-1} \left\{
\sum_{F_{\partial K}(2)}
 \int_{\sigma_2} \gamma^*
  ( \omega^2_{\a{0}{2} \a{1}{2}} ) \right. \\
&   & +  \left.
\sum_{F_{\partial K}(1)} 
 \int_{\sigma_1} \gamma^*
  (  \omega^1_{\a{0}{1} \a{0}{2} \a{1}{2}}
    -\omega^1_{\a{0}{1} \a{1}{1} \a{1}{2}})
\right\}  \\
&   & \times
\prod_{F_{\partial K}(0)} \left(
 g_{ \a{0}{0} \a{0}{1} \a{0}{2} \a{1}{2} }^{-1}
 g_{ \a{0}{0} \a{0}{1} \a{1}{1} \a{1}{2} }
 g_{ \a{0}{0} \a{1}{0} \a{1}{1} \a{1}{2} }^{-1}  \right)(\gamma(\sigma_0)) .
\end{eqnarray*}
\end{lem}

\begin{rem}
We omit the tilde for brevity.
\end{rem}

\begin{proof}
This is shown by direct computation using the cocycle condition of the \Cech cocycle and Stokes' theorem.
\end{proof}

\begin{lem} \label{lem_difference_of_action_2}
Let $\mu = (g,\omega^1,\omega^2,\omega^3), \mu' = (g',\omega'^1,\omega'^2,\omega'^3)$ be \Cech cocycle representations of $c_u$. We have $\mu'-\mu = D \nu$, where $\nu = (\xi,\pi^1,\pi^2)$ is a \Cech 2-cochain. When we define $\A_{M;A}$ using the cocycle representation $\mu$, we write $\A_{M,\mu;A}$. The following formula holds.
\begin{eqnarray*}
&   &
\A_{M,\mu+D\nu;A }(\gamma)
\A_{M,\mu;A}(\gamma)^{-1} \\
&   & =
\exp \sqrt{-1} \left\{
\sum_{F_{\partial K}(2)} \int_{\sigma^2} \gamma^* 
 \pi^2_{\a{}{2}}
-
\sum_{F_{\partial K}(1)} \int_{\sigma^1} \gamma^* 
 \pi^1_{\a{}{1} \a{}{2} }
\right\} \\
&   & \times
\prod_{F_{\partial K}(0)} 
 \xi_{\a{}{0} \a{}{1} \a{}{2} } (\gamma(\sigma_0)) .
\end{eqnarray*}
\end{lem}

\begin{proof}
The calculation of $\A_{M,D\nu;A }$ using Stokes' theorem gives the formula.
\end{proof}

\begin{thm}
Let $M$ be a compact oriented 3-manifold without boundary. We define a map
\begin{eqnarray*}
\A_M(\gamma) :  \M{M} \rightarrow \T
\end{eqnarray*}
by $\A_M|_{\W_A} = \A_{M;A}$ . This map is well-defined and depends only on the Deligne cohomology class $c_u$.
\end{thm}

\begin{proof}
The well-definedness of $\A_M$ is proved by Lemma \ref{lem_difference_of_action_1}. If we fix an open covering of $BG$, we can prove that $\A_M$ is independent of the cocycle representation by Lemma \ref{lem_difference_of_action_2}. In order to prove that $\A_M$ is independent of the choice of open covering of $BG$, we consider a refinement of an open covering. In this case we can take the induced covering of $\M{M}$ and cocycle representation of $c_u$ that does not change the value of $\A_M$. So this proposition is proved.
\end{proof}

We call $\A_M$ the CS action. The action satisfies the properties which are considered to be the axioms of ``Classical Field Theory." For the original action refer to \cite{F1}.

\begin{prop} \label{properties_of_action}
Let $M$ be a compact oriented 3-manifold without boundary. The CS action $\A_M$ satisfies the following properties.

(a)(Functoriality) Let $\phi^* : \M{M} \rightarrow \M{M'} $ be the map induced from a diffeomorphism $\phi : M' \rightarrow M$. For $\gamma \in \M{M}$, we have
$$
\A_{M'} (\phi^* \gamma) = \A_M (\gamma)  .
$$

(b)(Orientation) Let $M$ be a closed 3-manifold. We denote the manifold with opposite orientation to $M$ by $-M$. Then we have
$$
\A_{-M} (\gamma) = \overline{\A_M (\gamma)}  ,
$$
where the bar denotes complex conjugation.

(c)(Additivity) If $M = M_1 \sqcup M_2$ (disjoint union), then we denote the restriction of $\gamma$ to $M_1, M_2$ by $\gamma_1, \gamma_2$ respectively. We have
$$
\A_{M_1 \sqcup M_2} (\gamma) = 
\A_{M_1} (\gamma_1) \A_{M_2} (\gamma_2)  .
$$
\end{prop}

\begin{proof}
These properties are the consequences of the local sum definition of the action. For (a), we construct a triangulation $K'$ of $M'$ and a map $\alpha'$ by pulling back $K$ and $\alpha$ of $M$. Then the functoriality of the integration shows (a). If one reverses the orientation of $M$, then the integration over the oriented simplex changes sign. Thus (b) is proved. The formula in (c) is obtained by separating the summation of the simplices.
\end{proof}

\begin{prop} \label{prop_bounding_4mfd}
Let $\tilde{M}$ be a compact 4-manifold bounding $M$. If we have $\gamma : M \rightarrow BG$ by restricting $\tilde{\gamma} : \tilde{M} \rightarrow BG$, then we have
\begin{eqnarray*}
\A_M (\gamma) = 
\exp \frac{\sqrt{-1}}{4\pi}
\int_{\tilde{M}} \tilde{\gamma}^* Tr(F_u \wedge F_u)   .
\end{eqnarray*}
\end{prop}

\begin{proof}
If $(g, \omega^1, \omega^2, \omega^3)$ be a cocycle representation of $c_u$, then we have $\frac{1}{2\pi} d\omega^3_\alpha = \frac{1}{8\pi^2}Tr(F_u \wedge F_u) |_{U_\alpha}$ by definition. We triangulate $\tilde{M}$ by $\tilde{K} = \{ \tilde{\sigma}_0, \ldots, \tilde{\sigma}_4  \}$ and chose a map $\tilde{\phi} : \tilde{K} \rightarrow I$. We compute the right hand side of the formula as follows.
$$
\frac{\sqrt{-1}}{4\pi}
\int_{\tilde{M}} \tilde{\gamma}^* Tr(F_u \wedge F_u)
=
\sqrt{-1} \sum_{\tilde{\sigma}_4}
\int_{\tilde{\sigma}_4} \tilde{\gamma}^* 
d\omega^3_{\tilde{\phi}_{\tilde{\sigma}_4}}
=
\sqrt{-1} 
\sum_{ F_{\tilde{K}(3)} }
\int_{\tilde{\sigma}_3} \tilde{\gamma}^* 
\omega^3_{\tilde{\phi}_{\tilde{\sigma}_4}}   .
$$
Applying Stokes' theorem and the cocycle conditions, and canceling the summation over the simplices in the interior of $\tilde{M}$, we obtain the left hand side of the formula.
\end{proof}

The original Chern-Simons action is also computed by integrating over a bounding 4-manifold. This implies that our definition of the action for a general compact Lie group is an extension of the original one.

As stated in the introduction, Dijkgraaf and Witten \cite{D-W} formulated the CS action for general compact Lie groups using the differential characters of Cheeger-Simons. We chose the differential character $\alpha_u \in \hat{H}^3(BG,\R/\Z)$ specified by the characteristic 4-form $\frac{1}{8\pi^2}Tr(F_u \wedge F_u)$, where $F_u$ is the curvature form of the universal connection. In this case the action $S_M : \M{M} \rightarrow \R/\Z$ is defined by $S_M(\gamma) = \gamma^* \alpha_u(M)$. The isomorphism of Theorem \ref{thm3} and $H^3(BG) = 0$ imply the exact correspondence between $\alpha_u$ and $c_u$. This establishes the next result.

\begin{thm}
For an arbitrary closed 3-manifold, the formulation of the Chern-Simons action in this paper coincides with that of Dijkgraaf and Witten, i.e.\@ 
$\exp 2\pi\sqrt{-1}S_M(\gamma) = \A_M(\gamma)$.
\end{thm}

\medskip
%%%%%%%%%%%%%%%%%%%%%%%%%%%%%%%%

Next we formulate the CS action in the case that the underlying 3-manifold has boundary. For the purpose we describe a certain line bundle associated with a Riemann surface $\Sigma$ using Deligne cohomology. Let $K$ be a triangulation of $\Sigma$ and $\phi : K \rightarrow I$ be a map, where $I$ is the index set of an open covering of $BG$. Putting $A = \{K, \phi\}$ we have an open set $\U_A$ of $\M{\Sigma}$. The collection $\{ \U_A \}$ gives an open covering of $\M{\Sigma}$ (see Definition \ref{dfn_open_set}).

\begin{dfn} \label{dfn_cocycle_of_linebundle}
Let $\Sigma$ be a Riemann surface, $\{ U_{\alpha} \}$ an open covering of $BG$, and $(g, \omega^1, \omega^2, \omega^3)$ a \Cech cochain of $\check{C}^3(\{ U_\alpha \}, \F^3)$. We define a \Cech cochain $(G_{A_0 A_1}, \Omega_A) \in \check{C}^1(\{\U_A\}, \F^1)$ by the following
\begin{eqnarray*}
G_{A_0 A_1} (\gamma)
& = &
\exp \sqrt{-1} \left\{ \sum_{F_{K}(2)}
 \int_{\sigma_2} \gamma^*
  ( - \omega^2_{\a{0}{2} \a{1}{2}} )  \right. \\
& - & \left.
\sum_{F_{K}(1)} \int_{\sigma_1} \gamma^*
  (  \omega^1_{\a{0}{1} \a{0}{2} \a{1}{2}}
   - \omega^1_{\a{0}{1} \a{1}{1} \a{1}{2}}) \right\}  \\
& \times &
\prod_{F_{K}(0)} \left\{
 g_{ \a{0}{0} \a{0}{1} \a{0}{2} \a{1}{2}}
 g_{ \a{0}{0} \a{0}{1} \a{1}{1} \a{1}{2} }^{-1}
 g_{ \a{0}{0} \a{1}{0} \a{1}{1} \a{1}{2} }     \right\}(\gamma(\sigma_0)) , \\
%%%%%%%%%%%%%%%%%%%
\left( \iota_X \Omega_A \right)(\gamma) 
& = & -
\sum_{F_{K}(2)}  \int_{\sigma_2} \gamma^*
 \iota_X ( \omega^3_{\a{}{2}} )
-
\sum_{F_{K}(1)}  \int_{\sigma_1} \gamma^*
 \iota_X ( \omega^2_{\a{}{1} \a{}{2}} )   \\
& + &
\sum_{F_{K}(0)}
 \iota_X ( \omega^1_{ \a{}{0} \a{}{1} \a{}{2} } ) (\gamma(\sigma_0)) ,
\end{eqnarray*}
where $X$ is a section of $\gamma^*T(BG) \rightarrow \Sigma$ which is thought of as a tangent vector at $\gamma \in \M{\Sigma}$, and $\iota_X\omega$ is the inner product of a tangent vector $X$ and a differential form $\omega$.
\end{dfn}

\begin{lem} \label{closedness_of_cocycle_1}
If a Riemann surface $\Sigma$ is closed and $(g, \omega^1, \omega^2, \omega^3)$ is a cocycle, then the cochain $(G_{A_0 A_1}, \Omega_A)$ defined in Definition \ref{dfn_cocycle_of_linebundle} is a cocycle.
\end{lem}

\begin{proof}
We prove  $D(G_{A_0 A_1}, \Omega_A) = 0$. By using the cocycle condition of $(g, \omega^1, \omega^2, \omega^3)$ and Stokes' theorem, we have
\begin{eqnarray*}
(\delta G)_{A_0 A_1 A_2} (\gamma)
& = &
(G_{A_1 A_2} G^{-1}_{A_0 A_2} G_{A_0 A_1})(\gamma) \\
& =  & 
\exp \sqrt{-1} \left\{
 \sum_{F_{\partial K}(1)} \int_{\sigma_1} \gamma^*
   ( \omega^1_{ \a{0}{1} \a{1}{1} \a{2}{1} } ) \right\} \\
&    \times &
 \prod_{F_{\partial K}(0)} 
\left(
    g_{ \a{0}{0} \a{0}{1} \a{1}{1} \a{2}{1} }^{-1}
    g_{ \a{0}{0} \a{1}{0} \a{1}{1} \a{2}{1} }
    g_{ \a{0}{0} \a{1}{0} \a{2}{0} \a{2}{1} }^{-1} 
\right) (\gamma(\sigma_0)) 
\end{eqnarray*}
and
\begin{eqnarray*}
&   &
\iota_X \left(
(\delta \Omega)_{A_0 A_1} - \frac{1}{\sqrt{-1}}d \log G_{A_0 A_1}
\right)(\gamma)   \\
&   & =
\sum_{F_{\partial K}(1)} \int_{\sigma_1} \gamma^*
 \iota_X ( \omega^2_{ \a{0}{1} \a{1}{1} } )
-
\sum_{F_{\partial K}(0)} 
\left(
 \iota_X (  \omega^1_{ \a{0}{0} \a{0}{1}\a{1}{1} } 
          - \omega^1_{ \a{0}{0} \a{1}{0}\a{1}{1} } )
\right) (\gamma(\sigma_0)) .
\end{eqnarray*}
Because $\partial \Sigma = \emptyset$, the cocycle condition holds.
\end{proof}

Note that if a Riemann surface has boundary then the cochain $(G_{A_0 A_1}, \Omega_A)$ needs not be a cocycle.

\begin{cor} \label{lem_of_cech_hom}
If we define a homomorphism
\begin{eqnarray*}
\psi : \check{C}^3(\{ U_\alpha \},\F^3) \longrightarrow 
       \check{C}^1(\{ \U_A \},\F^1)
\end{eqnarray*}
by $\psi\left( ( g, \omega^1, \omega^2, \omega^3 ) \right) = (G, \Omega)$ using Definition \ref{dfn_cocycle_of_linebundle}, then we have 
$$
\psi \left( \check{Z}^3(\{ U_\alpha \},\F^3) \right) \subset \check{Z}^1(\{ \U_A \},\F^1) .
$$
\end{cor}

\begin{dfn}
Let $\Sigma$ be a Riemann surface without boundary. Fix a good covering of $BG$ and a \Cech cocycle representation of $c_u$. We have the open covering $\{ \U_{A} \}$ of $\M{\Sigma}$ and a cocycle $(G_{A_0 A_1}, \Omega_A)$. Then we define a Hermitian line bundle
\begin{eqnarray*}
\L_\Sigma  \longrightarrow \M{\Sigma} 
\end{eqnarray*}
by 
$$
\L_\Sigma = \coprod \left\{ \U_A \times \C \right\} / \sim ,
$$
where the equivalence relation is given by
$$
(\gamma, z_0 ) \sim (\gamma, z_1)
\Leftrightarrow 
z_0 = G_{A_0 A_1}(\gamma) z_1
$$ 
for $(\gamma, z_i) \in \U_{A_i} \times \C$. A Hermitian metric on $\L_\Sigma$ is defined by the usual metric on $\C$. We also define a Hermitian connection $\nabla$ by the 1-forms $\{ \Omega_A \}$. 
\end{dfn}

We call $\L_\Sigma$ the \textit{Chern-Simons line bundle}.

\begin{prop} \label{lem_independence_of_linebundle}
Let $\Sigma$ be a Riemann surface without boundary. The isomorphism class of the CS line bundle with the connection $(\L_\Sigma, \nabla)$ depends only on the Deligne cohomology class $c_u$.
\end{prop}

In order to prove this proposition we use the following theorem.

\begin{thm} \label{thm_transgression1}
The homomorphism of Corollary \ref{lem_of_cech_hom} induces the  homomorphism of Deligne cohomologies
\begin{eqnarray*}
\Psi : H^3(BG,\F^3) \longrightarrow 
       H^1(\M{\Sigma}, \F^1) .
\end{eqnarray*}
\end{thm}

\begin{proof}
First we fix an open covering $\{ U_\alpha \}$ and show that $\psi$ induces
$$
\check{\psi} : \check{H}^3(\{ U_\alpha \},\F^3) \longrightarrow 
               \check{H}^1(\{ \U_A \},\F^1) .
$$
For $( \xi, \pi^1, \pi^2 ) \in \check{C}^2(\{ U_\alpha \},\F^3)$ we define $(K_A) \in \check{C}^0(\{ \U_A \},\F^1)$ as follows.
\begin{eqnarray*}
K_A (\gamma) & = & 
\exp \sqrt{-1} 
\left\{ -
\sum_{F_{K}(2)} \int_{\sigma_2} \gamma^*
  \pi^2_{\a{}{2}}
+
\sum_{F_{K}(1)} \int_{\sigma_1} \gamma^*
  \pi^1_{\a{}{1} \a{}{2}} 
\right\}  \\
&   & \times
\prod_{F_{K}(0)} 
 \xi_{ \a{}{0} \a{}{1} \a{}{2} }^{-1} (\gamma(\sigma_0)) .
\end{eqnarray*}
We can verify $\psi \left( D(\xi, \pi^1, \pi^2) \right) = D(K_A)$ by direct computation using Stokes' theorem. This implies that
$$
\psi \left( \check{B}^3(\{ U_\alpha \},\F^3) \right) \subset \check{B}^1(\{ \U_A \},\F^1) .
$$
Hence $\psi$ induces the homomorphism $\check{\psi}$ of \Cech cohomologies. 

Secondly, we show that the homomorphism of Deligne cohomologies induced from $\check{\psi}$ is well-defined. An open covering $\{ U'_{\alpha'} \}$ is a refinement of $\{ U_\alpha \}$ if there exists a map of indices $\rho : \{\alpha'\} \rightarrow \{\alpha \}$ such that $U'_{\alpha'} \subset U_{\rho(\alpha')}$. The refinement of the open covering of $BG$ induces a refinement of the coverings of the map space as follows. If a triangulation $K$ of $\Sigma$ and a map $\phi' : K \rightarrow \{\alpha'\}$ are given, then we have an open set
$$
\U'_{A'} = \{\gamma \in \M{\Sigma} |
\gamma(\sigma_p) \subset U'_{\phi'_{\sigma_p}} (p=0,1,2) \} ,
$$
where $A'=\{K, \phi'\}$. We can define a map $R : \{A'\} \rightarrow \{A\}$ of indices of the coverings of the map space by $R(A') = \{K, \rho \circ \phi'\}$. It is easy to see that $\U'_{A'} \subset \U_{R(A')}$. So we have a refinement $\{ \U'_{A'} \}$ of $\{ \U_A \}$. These refinements induce homomorphisms  of \Cech cohomologies $\check{\rho},\check{R}$ which commute with $\check{\psi}$ as follows.
$$
\begin{CD}
\check{H}^3(\{ U_\alpha \},\F^3) @>\check{\rho}>>
\check{H}^3(\{ U'_{\alpha'} \},\F^3) \\
@V\check{\psi}VV  @VV\check{\psi}V  \\
\check{H}^3(\{ \U_A \},\F^3) @>\check{R}>>
\check{H}^3(\{ \U'_{A'} \},\F^3)
\end{CD}
$$
Taking the direct limit, we obtain the well-defined homomorphism
$$
\Psi : H^3( BG, \F^3) \longrightarrow 
       H^1( \M{\Sigma}, \F^1) ,
$$
which is induced from $\check{\psi}$.
\end{proof}

\begin{rem}
The general formula of Theorem \ref{thm_transgression1} is known \cite{G-T2} .
\end{rem}

\begin{proof}[Proof of Proposition \ref{lem_independence_of_linebundle}]
By Theorem \ref{thm2} an isomorphism class of a line bundle with a connection corresponds uniquely to a Deligne cohomology class. The Deligne cohomology class that corresponds to the isomorphism class of $(\L_\Sigma, \nabla)$ is expressed by $\Psi(c_u)$. So the isomorphism class is uniquely determined by $c_u$.
\end{proof}

\begin{prop}
The curvature form of $(\L_\Sigma, \nabla)$ is
\begin{eqnarray*}
\frac{1}{4\pi} \int_\Sigma ev^* Tr(F_u \wedge F_u) ,
\end{eqnarray*}
where $ev : \M{\Sigma} \times \Sigma \rightarrow BG$ is the evaluation map and $\int_\Sigma$ is the fiber integration along $\Sigma$.
\end{prop}

\begin{proof}
Let $(g, \omega^1, \omega^2, \omega^3)$ be a cocycle representation of $c_u$. Note that $\frac{1}{2\pi} d\omega^3_\alpha = \frac{1}{8\pi^2}Tr(F_u \wedge F_u) |_{U_\alpha}$. We obtain the curvature form by the computation of $-d\Omega_A$ using the following formula
$$
\iota_{X_k} \cdots \iota_{X_0} d \int_{\sigma_p} \gamma^* \omega^{k+p} 
=
\int_{\sigma_p} \gamma^*
(          \iota_{X_k} \cdots \iota_{X_0} d
  + (-1)^k d \iota_{X_k} \cdots \iota_{X_0} )
\omega^{k+p} .
$$
\end{proof}

For a 3-manifold $M$ with boundary $\Sigma$ we denote the restriction map by
$$
r : \M{M} \rightarrow \M{\Sigma} .
$$
The pull back $r^*\L_\Sigma$ is by definition $r^*\L_\Sigma = \coprod \{ r^{-1}\U_A \times \C \} / \sim $. It is easy to see that $r^{-1}\U_A = \bigcup \W_{\tilde{A}}$, where the open set $\W_{\tilde{A}}$ with $\tilde{A} = \{ \tilde{K}, \tilde{\phi} \}$ is defined by the triangulation $\tilde{K}$ of $M$ such that $\partial \tilde{K} = K$, and the map $\tilde{\phi} : \tilde{K} \rightarrow I$ such that $\tilde{\phi}|_{\partial \tilde{K}} = \phi$. When $A=\{K, \phi\}$ and $\tilde{A} = \{ \tilde{K}, \tilde{\phi} \}$ have the above relation, we write $\partial \tilde{A} = A$ for short. 

\begin{thm}
Let $M$ be a 3-manifold with $\partial M = \Sigma$, and fix a cocycle representation of $c_u$. We define a section 
\begin{eqnarray*}
\A_M : \M{M} \rightarrow r^*\L_\Sigma
\end{eqnarray*}
by $\A_M|_{r^{-1}\U_A}(\gamma) = (\gamma, \A_{M;\tilde{A}}(\gamma))$, where $\partial \tilde{A} = A$. This section is well-defined and takes its values in the unit norm.
\end{thm}

\begin{proof}
Let $\W_{\tilde{A}_0}, \W_{\tilde{A}_1}$ be two open sets such that $\partial \tilde{A}_0 = \partial \tilde{A}_1 = A$. By Lemma \ref{lem_difference_of_action_1} we have $\A_{M;\tilde{A_0}} = \A_{M;\tilde{A_1}}$. So $\A_M|_{r^{-1}\U_A}$ is well-defined. Let $\U_{A_0}, \U_{A_1}$ be open sets of $\M{\Sigma}$. On the non-empty intersection $r^{-1}\U_{A_0} \cap r^{-1}\U_{A_1}$ we have $\A_M|_{r^{-1}\U_{A_0}} = G_{A_0 A_1} \A_M|_{r^{-1}\U_{A_1}}$ using again Lemma \ref{lem_difference_of_action_1}. This shows that $\A_M$ is indeed the section of $r^*\L_\Sigma$. By definition $\A_{M;\tilde{A}}$ takes its values in the unit norm. 
\end{proof}

We also call $\A_M$ the CS action.

\begin{prop}
Let $M$ be a 3-manifold with boundary $\Sigma$, and $\mu, \mu'$ cocycle representations of $c_u$. When we use the cocycle representation $\mu$ to define the CS action and the CS line bundle, we write $A_{M,\mu}$ and $\L_{\Sigma,\mu,}$. The CS action satisfies the following formula
$$
\phi_*(\A_{M,\mu}) = \A_{M,\mu'},
$$
where $\phi_*$ is the linear map on the space of sections induced from the bundle isomorphism $\phi : \L_{\Sigma, \mu} \rightarrow \L_{\Sigma, \mu'}$.
\end{prop}

\begin{proof}
If two cocycle representations $\mu, \mu'$ are given, then we have $\mu'-\mu = D\nu$, where $\nu = (\xi, \pi^1, \pi^2)$ . By Proposition \ref{lem_independence_of_linebundle}, the line bundles $\L_{\Sigma, \mu}$ and $\L_{\Sigma, \mu'}$ are isomorphic. The bundle isomorphism $\phi$ is defined by
$$
(\gamma, z) \mapsto (\gamma, z K_A(\gamma)) ,
$$
where $(\gamma, z) \in \U_A \times \C$ and $K_A$ is the function defined in the proof of Theorem \ref{thm_transgression1}. By Lemma \ref{lem_difference_of_action_2} we have $\phi_*(\A_{M,\mu} |_{r^{-1}\U_A}) = A_{M,\mu'} |_{r^{-1}\U_A}$. This implies the formula in the proposition.
\end{proof}

For $\gamma \in \M{M}$ we canonically identify the fiber $r^* \L_{\partial M} \big|_\gamma$ with $\L_{\partial M} \big|_{ \partial \gamma}$, where $\partial \gamma = r(\gamma)$. So the CS action takes its values in the fiber of the CS line bundle $\A_M(\gamma) \in \L_{\partial M} \big|_{\partial \gamma}$.

\begin{prop} \label{prop_line}
Let $M$ be a compact oriented 3-manifold with non-empty boundary $\partial M = \Sigma$. The value of the CS action and the CS line bundle $\L_\Sigma$ satisfy the following properties.

(a)(Functoriality) Let $\phi^* : \M{\Sigma} \rightarrow \M{\Sigma'}$ be the map induced from a diffeomorphism $\phi : \Sigma' \rightarrow \Sigma$. For $\gamma \in \M{M}$ there is a natural isometry
$$
\overline{\phi^*}  : 
\L_{\Sigma} \big|_{\partial \gamma} 
\stackrel{\sim}{\rightarrow} 
\L_{\Sigma'} \big|_{\phi^* (\partial \gamma)} .
$$
Moreover, if $\phi : \Sigma' \rightarrow \Sigma$ is the restriction of $\Phi : M' \rightarrow M$, then we have
$$
\A_{M'} (\Phi^* (\gamma)) = \overline{\phi^*} ( \A_M (\gamma) )  .
$$

(b)(Orientation) Concerning the orientations of manifolds, the following holds.
\begin{eqnarray*}
 \L_{-\Sigma}    \big|_{\partial \gamma} & \cong & 
{\L_{ \Sigma}}^* \big|_{\partial \gamma}, \\
%%%
          \A_{-M}(\gamma)   & = &
\overline{\A_{ M}(\gamma)}.
\end{eqnarray*}

(c)(Additivity) If $\Sigma = \Sigma_1 \sqcup \Sigma_2 $, then we have
$$
\L_{\Sigma_1 \sqcup \Sigma_2} 
 \big|_{\partial \gamma_1 \sqcup \partial \gamma_2} 
\cong 
\L_{\Sigma_1} \big|_{\partial \gamma_1} \otimes 
\L_{\Sigma_2} \big|_{\partial \gamma_2}  .
$$
Moreover if $M = M_1 \sqcup M_2$, with $\partial M_i = \Sigma_i$, then we have
$$
\A_{M_1 \sqcup M_2} (\gamma_1 \sqcup \gamma_2) =
\A_{M_1} (\gamma_1) \otimes \A_{M_2} (\gamma_2)
$$
under the isometry.

(d)(Gluing) Suppose that a closed Riemann surface $\Sigma$ is embedded in $M$. If we cut $M$ along $\Sigma$, we have a new manifold $M_{cut}$ with $\partial M_{cut} = \partial M \sqcup \Sigma \sqcup -\Sigma$. Let $\gamma_{cut}$ be the map induced from $\gamma$ by restricting $M$ to $M_{cut}$, and $\xi$ the map induced from $\gamma$ by restricting $M$ to $\Sigma$. Then there is the contraction
$$
Tr_{\xi} :
\L_{\partial M_{cut}} \big|_{\partial \gamma_{cut}} 
\cong 
\L  _{\partial M}     \big|_{\partial \gamma} \otimes 
\L  _{\Sigma}         \big|_{\xi}             \otimes 
\L^*_{\Sigma}         \big|_{\xi}
\rightarrow
\L_{\partial M} \big|_{\partial \gamma}  ,
$$
and we have
$$
Tr_{\xi}( \A_{M_{cut}} (\gamma_{cut}) ) =
 \A_M (\gamma)  .
$$
\end{prop}

\begin{proof}
Note that the CS action and the transition functions of the line bundle are expressed by summations of local terms. So the properties (a), (b), (c), (d) are verified in the proof of Proposition \ref{properties_of_action}.
\end{proof}

\begin{prop}
Let $\Sigma$ be a Riemann surface without boundary and $I$ the unit interval. We identify a map $\Gamma : \Sigma \times I \rightarrow BG$ with a path $\gamma : I \rightarrow \M{\Sigma}$. Then the value of the CS action
\begin{eqnarray*}
\A^{-1}_{\Sigma \times I} (\Gamma) 
\in
\L_{\Sigma}^* \big|_{\gamma_0} \otimes
\L_{\Sigma}   \big|_{\gamma_1}
\end{eqnarray*}
coincides with the parallel transport along $\gamma$ determined by the connection $\nabla$.
\end{prop}

\begin{proof}[Sketch of the proof]
We should verify that the CS action of $\Gamma$ satisfies the differential equation of the parallel transport determined by the connection $\nabla$. For this purpose we construct a path in $\U_A$ by endowing $\Sigma \times I$ with a specific triangulation that is induced from the triangulation of $\Sigma$. Then explicit calculation of the action with respect to this triangulation shows the following
\begin{eqnarray*}
\frac{d}{dt} \left( 
 \frac{1}{\sqrt{-1}} \log 
  \A_{ \Sigma \times [0,t] } 
   ( \Gamma \big|_{ \Sigma \times [0,t] } ) \right)
= - \iota _{\dot{\gamma}} \Omega_{A} .
\end{eqnarray*}
This will complete the proof.
\end{proof}

\begin{prop}
A map $\gamma : M \rightarrow BG$ is a stationary point of the action $\A_M$ if and only if it induces a flat bundle.
\end{prop}

\begin{proof}
First we take a map $\gamma _0 : M \rightarrow BG$. We can make a path of connections on $\gamma_0^*EG$ by setting $A_t = \gamma _0 ^* A_u + t \alpha$, where $\alpha$ is a $\g$-valued 1-form on $\gamma_0^*EG$ and $\alpha \big|_{\partial M} = 0$ if $\partial M \neq \emptyset$. This is also a connection on the bundle over $M \times I$, so we take $\gamma _t : M \times I \rightarrow BG$ correspondingly. Using Proposition \ref{prop_bounding_4mfd} we have
$$
\A_M (\gamma _t) \A_M (\gamma_0)^{-1}
=\A_{\partial M \times [0,t]}(\gamma_t \big|_{\partial M})^{-1}
\exp \frac{\sqrt{-1}}{4 \pi}
 \int_{M \times [0,t]} {\gamma _t}^* Tr( F_u \wedge F_u)
$$
and $\A_{\partial M} (\gamma_t \big|_{\partial M})$ is independent of $t$. We denote the curvature of $A_t$ on the bundle over $M \times \{t\}$ by $F_t$. The curvature of $A_t$ on the bundle over $M \times I$ is $F_t+dt\wedge\alpha$. Finally we have
$$
{\gamma _t}^* Tr ( F_u \wedge F_u )
= 2 dt \wedge Tr ( \alpha \wedge F_t) .
$$
Hence the connection $A_0$ induced by $\gamma_0$ is a stationary point of $\A_M$ if and only if $F_0 = 0$.
\end{proof}

%%%%%%%%%%%%%%%%%%%%%%%%%%%%%%%%%%%%%%%%%%%

Graduate school of Mathematical Sciences, 
University of Tokyo, Komaba 3-8-1, Meguro-Ku, Tokyo, 153-8914 Japan.

e-mail: kgomi@ms.u-tokyo.ac.jp

\end{document}